%% file: main.tex
\def\ispreprint{} 

\def\isfinal{}

\ifdefined\isiclrpaper
    \documentclass{article}
\else\ifdefined\ispreprint
    \documentclass[10pt]{article}
\fi
\fi

\input{preamble}

\usetikzlibrary{external}
\begin{document}


\maketitle

\draftnotice
\conditionalLoT 
\conditionalLoE 

\begin{abstract}
Implicit neural representations (INRs) have arisen as useful methods for representing signals on Euclidean domains. 
By parameterizing an image as a multilayer perceptron (MLP) on Euclidean space, INRs effectively represent signals in a way that couples spatial and spectral features of the signal that is not obvious in the usual discrete representation, paving the way for continuous signal processing and machine learning approaches that were not previously possible. 
Although INRs using sinusoidal activation functions have been studied in terms of Fourier theory, recent works have shown the advantage of using wavelets instead of sinusoids as activation functions, due to their ability to simultaneously localize in both frequency and space. 
In this work, we approach such INRs and demonstrate how they resolve high-frequency features of signals from coarse approximations done in the first layer of the MLP. 
This leads to multiple prescriptions for the design of INR architectures, including the use of complex wavelets, decoupling of low and band-pass approximations, and initialization schemes based on the singularities of the desired signal.
\end{abstract}

\section{Introduction}
\label{sec:intro}

Implicit neural representations (INRs) have emerged as a set of neural architectures for representing and processing signals on low-dimensional spaces.
By learning a continuous interpolant of a set of sampled points, INRs have enabled and advanced state-of-the-art methods in signal processing~\citep{xu2022} and computer vision~\citep{mildenhall2020}.

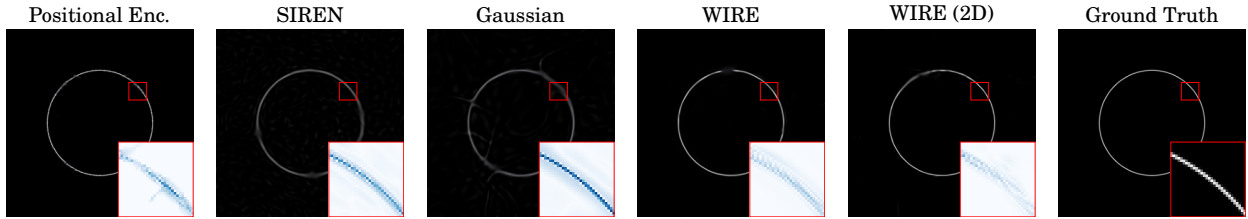
\begin{figure*}
  \centering
  \resizebox{\linewidth}{!}{\input{figs/zoo/zoo.tikz}}
  \caption{Zoo of INRs for representing a simple image $\mathbb{R}^2\to\mathbb{R}$. Highlighted squares indicate error relative to true image.}
  \label{fig:zoo}
\end{figure*}

Typical INRs are specially designed multilayer perceptrons (MLPs), where the activation functions are chosen in such a way to yield a desirable signal representation; some of these methods are demonstrated on a simple test image in \cref{fig:zoo}.

Although these INRs can be easily understood at the first layer due to the simplicity of plotting the function associated to each neuron based on its weights and biases, the behavior of the network in the second layer and beyond is more opaque, apart from some theoretical developments in the particular case of a sinusoidal first layer~\citep{yuce2022}.

This work develops a broader theoretical understanding of INR architectures with a wider class of activation functions, followed by practical prescriptions rooted in time-frequency analysis.
In particular, we
\begin{enumerate}
    \item Characterize the function class of INRs in terms of Fourier convolutions of the neurons in the first layer (\cref{thm:inr})
    \item Demonstrate how INRs that use complex wavelet functions preserve useful properties of the wavelet, even after the application of the nonlinearities (\cref{coro:prog})
    \item Suggest a split architecture for approximating signals, decoupling the smooth and nonsmooth parts into linear and nonlinear INRs, respectively (\cref{sec:complex:split})
    \item Leverage connections with wavelet theory to propose efficient initialization schemes for wavelet INRs based on the wavelet modulus maxima for capturing singularities in the target functions (\cref{sec:singular}).
\end{enumerate}
Following a brief survey of INR methods, the class of architectures we study is defined in \cref{sec:survey}.
The main result bounding the function class represented by these architectures is stated in \cref{sec:bound}, which is then related to the algebra of complex wavelets in \cref{sec:complex}.
The use of the wavelet modulus maxima for initialization of wavelet INRs is described and demonstrated in \cref{sec:singular}, before concluding in \cref{sec:conclusion}.

\section{Implicit Neural Representations}
\label{sec:survey}

Wavelets as activation functions in MLPs have been shown to yield good function approximators~\citep{zhang1992,marar1996}.
These works have leveraged the sparse representation of functions by wavelet dictionaries in order to construct simple neural architectures and training algorithms for effective signal representation.
Indeed, an approximation of a signal by a finite linear combination of ridgelets~\citep{candes1998} can be viewed as one such MLP using wavelet activation functions.

Recently, sinusoidal activation functions in the first layer~\citep{tancik2020} and beyond~\citep{sitzmann2020,fathony2020} have been shown to yield good function approximators, coupled with a harmonic analysis-type bound on the function class represented by these networks~\citep{yuce2022}.
Other methods have used activation functions that, unlike sinusoids, are localized in space, such as gaussians~\citep{ramasinghe2021} or Gabor wavelets~\citep{saragadam2023}.

Following the formulation of \citep{yuce2022}, we define an INR to be a map $f_{\boldsymbol{\theta}}:\mathbb{R}^d\to\mathbb{C}$ defined in terms of a function $\psi:\mathbb{R}^d\to\mathbb{C}$, followed by an MLP with analytic\footnote{That is, entire on $\mathbb{C}$.} activation functions $\rho^{(\ell)}:\mathbb{C}\to\mathbb{C}$ for layers $\ell=1,\ldots,L$:
\begin{equation}\label{eq:inr}
  \begin{aligned}
    \mathbf{z}^{(0)}(\mathbf{r}) &= \psi(\mathbf{W}^{(0)}\mathbf{r} + \mathbf{b}^{(0)}) \\
    \mathbf{z}^{(\ell)}(\mathbf{r}) &= \rho^{(\ell)}(\mathbf{W}^{(\ell)}\mathbf{z}^{(\ell-1)}(\mathbf{r}) + \mathbf{b}^{(\ell)}) \\
    f_{\boldsymbol{\theta}}(\mathbf{r}) &= \mathbf{W}^{(L)}\mathbf{z}^{(L-1)}(\mathbf{r}) + \mathbf{b}^{(L)},
  \end{aligned}
\end{equation}
where $\boldsymbol{\theta}$ denotes the set of parameters dictating the tensor $\mathbf{W}^{(0)}\in\mathbb{R}^{F_1\times d\times d}$, matrices $\mathbf{W}^{(\ell)}\in\mathbb{C}^{F_{\ell+1}\times F_{\ell}}$ and $\mathbf{b}^{(0)}\in\mathbb{R}^{F_1\times d}$, and vectors $\mathbf{b}^{(\ell)}\in\mathbb{C}^{F_{\ell+1}}$ for $\ell=1,\ldots,L$, with fixed integers $F_\ell$ satisfying $F_{L+1}=1$.
We will henceforth refer to $\psi$ as the \emph{template function} of the INR.
Owing to the use of Gabor wavelets by \citet{saragadam2023}, we will refer to functions of the form~\eqref{eq:inr} as \emph{WIRE INRs}, although~\eqref{eq:inr} also captures architectures that do not use wavelets, such as SIREN~\citep{sitzmann2020}.

\section{Main Results}
\label{sec:bound}

For the application of INRs to practical problems, it is important to understand the function class that an INR architecture can represent.
We will demonstrate how the function parameterized by an INR can be understood via time-frequency analysis, ultimately motivating the use of wavelets as template functions.

\subsection{Expressivity of INRs}

Noting that polynomials of sinusoids generate linear combinations of integer harmonics of said sinusoids, \citet{yuce2022} bounded the expressivity of SIREN~\citep{sitzmann2020} and related architectures~\citep{fathony2020}.
These results essentially followed from identities relating products of trigonometric functions.
For template functions that are not sinusoids, such as wavelets~\citep{saragadam2023}, these identities do not hold.
The following result offers a bound on the class of functions represented by an INR.
\begin{theorem}\label{thm:inr}
  Let $f_{\boldsymbol{\theta}}:\mathbb{R}^d\to\mathbb{C}$ be a WIRE INR.
  Assume that each of the activation functions $\rho^{(\ell)}$ is a polynomial of degree at most $K$, and that the Fourier transform of the template function $\psi$ exists.\footnote{It may exist only in the sense of tempered distributions.}
  Let $\mathbf{W}^{(0)}\mathbf{r}=[\mathbf{W}_1\mathbf{r},\ldots,\mathbf{W}_{F_1}\mathbf{r}]^\top$ for $\mathbf{W}_1,\ldots,\mathbf{W}_{F_1}\in\mathbb{R}^{d\times d}$ each having full rank, and also let $\mathbf{b}^{(0)}=[\mathbf{b}_1,\ldots,\mathbf{b}_{F_1}]^\top$ for $\mathbf{b}_1,\ldots,\mathbf{b}_{F_1}\in\mathbb{R}^d$.
  For $k\geq 0$, denote by $\Delta(F_1,k)$ the set of ordered $F_1$-tuples of nonnegative integers $\boldsymbol{\ell}=[\ell_1,\ldots,\ell_{F_1}]$ such that $\sum_{t=1}^{F_1}\ell_t=k$.

  Let a point $\mathbf{r}_0\in\mathbb{R}^d$ be given.
  Then, there exists an open neighborhood $U\ni\mathbf{r}_0$ such that for all $\phi\in\mathcal{C}_0^\infty(U)$
  \begin{equation}\label{eq:fourier-rep}
    \widehat{\phi\cdot f_{\boldsymbol{\theta}}}(\xi) 
    = 
    \left( \widehat{\phi} \ast \sum_{k=0}^{K^{L-1}} \sum_{\boldsymbol{\ell}\in\Delta(F_1,k)} \widehat{\beta}_{\boldsymbol{\ell}} \bigast_{t=1}^{F_1} \left( e^{i2\pi\langle\mathbf{W}_t^{-\top}\xi,\mathbf{b}_t\rangle} \widehat{\psi}(\mathbf{W}_t^{-\top}\xi) \right)^{\ast \ell_t, \xi} \right) (\xi),
  \end{equation}
  for coefficients $\widehat{\beta}_{\boldsymbol{\ell}}\in\mathbb{C}$ independent of $\mathbf{r}$, where $(\cdot)^{\ast \ell, \xi}$ denotes $\ell$-fold convolution\footnote{$0$-fold convolution is defined by convention to yield the Dirac delta.} of the argument with itself with respect to $\xi$.
  Furthermore, the coefficients $\widehat{\beta}_{\boldsymbol{\ell}}$ are only nonzero when each $t\in[1,\ldots,F_1]$ such that $\ell_t\neq 0$ also satisfies $\mathbf{W}_t\mathbf{r}_0+\mathbf{b}_t\in\supp(\psi)$.
\end{theorem}
The proof is left to \cref{app:inr}.
\Cref{thm:inr} illustrates two things.
First, the output of an INR has a Fourier transform determined by convolutions of the Fourier transforms of the atoms in the first layer with themselves, serving to generate ``integer harmonics'' of the initial atoms determined by scaled, shifted copies of the template function $\psi$.
Notably, this recovers~\citep[Theorem~1]{yuce2022}.
Second, the support of these scaled and shifted atoms is preserved, so that the output at a given coordinate $\mathbf{r}$ is dependent only upon the atoms in the first layer whose support contains $\mathbf{r}$.

\begin{remark}
  The assumptions behind \cref{thm:inr} can be relaxed to capture a broader class of architectures.
  By imposing continuity conditions on the template function $\psi$, the activation functions can be reasonably extended to analytic functions.
  These extensions are discussed in \cref{app:templates}.
\end{remark}

\subsection{Effective Time-Frequency Support of INRs}

As noted before, \cref{thm:inr} describes the output of an INR at a point by self-convolutions of sums of the Fourier transforms of the atoms in the first layer whose support contains that point.
So, for the remainder of this section, we can assume that $\mathbf{r}_0\in\mathbb{R}^d$ is fixed, and that we only consider indices $t$ such that $\psi(\mathbf{W}_t\mathbf{r}_0+\mathbf{b}_t)\neq 0$.

Assume that the template function $\psi$ has compact support, which precludes $\widehat{\psi}$ from having compact support.
However, if we make the further assumption that $\psi$ is smooth, then $\widehat{\psi}$ is rapidly decreasing.
We will then say, informally, that $\widehat{\psi}$ has a compact \emph{effective} support, denoted by the set $A\subset\mathbb{R}^d$.
Making the approximation $\supp(\widehat{\psi})\approx A$, we have that $\supp(\widehat{\psi}(\mathbf{W}_t^{-\top}\xi))\approx\mathbf{W}_t^\top A$.

As a first approximation, then, we can estimate the frequency support ``locally'' around the point $\mathbf{r}_0\in\mathbb{R}^d$, in the sense of the Fourier transform following multiplication with a suitable cutoff function.
Examining the summands in \eqref{eq:fourier-rep}, the support of the self-convolutions in the Fourier domain are bounded by
\begin{equation*}
  \supp \left( \bigast_{t=1}^{F_1} \left( e^{i2\pi\langle\mathbf{W}_t^{-\top}\xi,\mathbf{b}_t\rangle} \widehat{\psi}(\mathbf{W}_t^{-\top}\xi) \right)^{\ast \ell_t, \xi} \right)
  \subset \sum_{t=1}^{F_1} \ell_t\mathbf{W}^\top A,
\end{equation*}
where the subset relationship is understood to be in the informal sense of effective support, and the sum on the RHS is understood as the Minkowski sum of the summands $\ell_t\mathbf{W}_t^\top A$.
So, the frequency support described by \cref{thm:inr} is effectively bounded by the set
\begin{equation*}
  \bigcup_{k=0}^{K^{L-1}}\left\{\sum_{t=1}^{F_1} \ell_t\mathbf{W}_t^\top A: \ell_t\in\Delta(F_1,k)\right\}.
\end{equation*}

\section{The Algebra of Complex Wavelets}
\label{sec:complex}

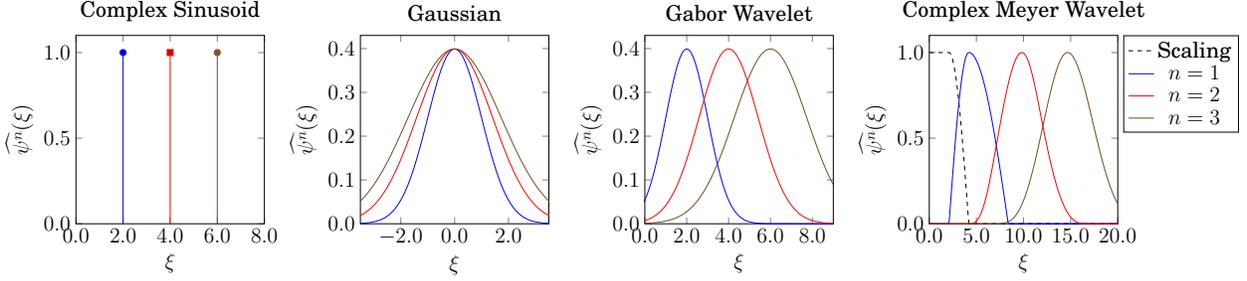
\begin{figure*}
    \centering
    \resizebox{\linewidth}{!}{\input{figs/stable/stable.tikz}}
    \caption{Template functions in Fourier domain under exponentiation.}
    \label{fig:stable}
\end{figure*}

Of the INR architectures surveyed in \cref{sec:survey}, the only one to use a complex wavelet template function is WIRE~\citep{saragadam2023}, where a Gabor wavelet is used. 
Gabor wavelets are essentially asymmetric (in the Fourier domain) band-pass filters, which are necessarily complex-valued due to the lack of conjugate symmetry in the Fourier domain.
We now consider the advantages of using complex wavelets, or more precisely \emph{progressive wavelets}, as template functions for INRs by examining their structure as an algebra of functions.

\subsection{Progressive Template Functions}

For the sake of discussion, suppose that $d=1$, so that the INR represents a 1D function.
The template function $\psi:\mathbb{R}\to\mathbb{C}$ is said to be \emph{progressive}\footnote{More commonly known as an \emph{analytic signal}, we use this terminology to avoid confusion with the analytic activation functions used in the INR.}
if it has no negative frequency components, \ie, for $\xi<0$, we have $\widehat{\psi}(\xi)=0$~\citep{mallat1999}.
If $\psi$ is integrable, its Fourier transform is uniformly continuous, which implies that for integrable progressive functions we have $\widehat{\psi}(0) = 0$.
Without embarking upon a long review of progressive wavelets~\citep{mallat1999},
we discuss some of the basic properties of progressive functions that are relevant to the discussion at hand.
It is obvious that progressive functions remain progressive under scalar multiplication, shifts, and positive scaling.
That is, for arbitrary $s> 0,u\in\mathbb{R},z\in\mathbb{C}$, if $\psi(x)$ is progressive, then the function $z\cdot D_s T_u \psi(x) := z\cdot\psi((x-u)/s)$ is also progressive.
Moreover, progressive functions are closed under multiplication, so that if $\psi_1$ and $\psi_2$ are progressive, then $\psi_3(x):=\psi_1(x)\psi_2(x)$ is also progressive,\footnote{This is a simple consequence of the convolution theorem.} \ie, progressive functions constitute an algebra over $\mathbb{C}$.

\begin{example}[Complex Sinusoid]
    For any $\omega>0$, the complex sinusoid $\psi(x;\omega)=\exp(-i2\pi\omega x)$ is a progressive function, as its Fourier transform is a Dirac delta centered at $\omega$.
    As pictured in \cref{fig:stable}~(Left), the exponents $\psi^n(\cdot;\omega)$ are themselves complex sinusoids, where $\psi^n(x;\omega)=\psi(x;n\omega)$.
\end{example}

\begin{example}[Gaussian]
    The gaussian function, defined for some $\sigma>0$ as $\psi(x;\sigma)=\exp(-x^2/(2\sigma^2))$, is \emph{not} a progressive function, as its Fourier transform is symmetric and centered about zero.
    Moreover, as pictured in \cref{fig:stable}~(Center-Left), the exponents are also gaussian functions $\psi^n(x;\sigma)=\psi(x;\sigma/\sqrt{n})$, which also have Fourier transform centered at zero.
    Unlike the complex sinusoid, the powers of the gaussian are all low-pass, but with increasingly wide passband.
\end{example}

\begin{example}[Gabor Wavelet]
    For any $\omega,\sigma>0$, the Gabor wavelet defined as $\psi(x;\omega,\sigma)=\exp(-x^2/(2\sigma^2)-i2\pi\omega x)$ is \emph{not} a progressive function, as its Fourier transform is a gaussian centered at $\omega$ with standard deviation $1/\sigma$.
    However, the Fourier transform of the exponents $\psi^n$ for integers $n>0$ are gaussians centered at $n\omega$ with standard deviation $\sqrt{n}/\sigma$, as pictured in \cref{fig:stable}~(Center-Right).
    So, as $n$ grows sufficiently large, the effective support of $\widehat{\psi^n}$ will be contained in the positive reals, so that the Gabor wavelet can be considered as a progressive function for the purposes of studying INRs.
\end{example}

A progressive function on $\mathbb{R}$ has Fourier support contained in the nonnegative real numbers.
Of course, there is not an obvious notion of nonnegativity that generalizes to $\mathbb{R}^d$ for $d>1$.
Noting that the nonnegative reals form a convex conic subset of $\mathbb{R}$, we define the notion of a progressive function with respect to some conic subset of $\mathbb{R}^d$:
\begin{definition}\label{defn:progressive}
    Let $\Gamma\subseteq\widehat{\mathbb{R}}^d$ be a convex conic set, \ie, for all $\gamma_1,\gamma_2\in\Gamma$ and $a_1,a_2\geq 0$, we have that $a_1\gamma_1+a_2\gamma_2\in\Gamma$.\footnote{We henceforth refer to such sets as simply ``conic.''}
    A function $\psi:\mathbb{R}^d\to\mathbb{C}$ is said to be \emph{$\Gamma$-progressive} if $\supp(\widehat{\psi})\subseteq\Gamma$.
    The function $\psi$ is said to be \emph{locally $\Gamma$-progressive at $\mathbf{r}_0\in\mathbb{R}^d$} if there exists some $\Gamma$-progressive function $\psi_{\mathbf{r}_0}:\mathbb{R}^d\to\mathbb{C}$ so that for all smooth functions $\phi\in\mathcal{C}_0^\infty(\mathbb{R}^d)$ with support in a sufficiently small neighborhood of $\mathbf{r}$, we have
    \begin{equation}
        \widehat{\phi\cdot\psi} = \widehat{\phi}\ast\widehat{\psi_{\mathbf{r}_0}}.
    \end{equation}
\end{definition}
Curvelets~\citep{candes2004}, for instance, are typically defined in a way to make them $\Gamma$-progressive for some conic set $\Gamma$ that indicates the oscillatory direction of a curvelet atom.
Observe that if $\Gamma$ is a conic set, then for any matrix $\mathbf{W}$, the set $\mathbf{W}^\top\Gamma$ is also conic.
Thus, for a function $\psi$ that is $\Gamma$-progressive, the function $\psi(\mathbf{W}\mathbf{x})$ is $\mathbf{W}^{\top}\Gamma$-progressive.

Observe further that for two $\Gamma$-progressive functions $\psi_1,\psi_2$, their product is also $\Gamma$-progressive.
The closure of progressive functions under multiplication implies that an analytic function applied pointwise to a progressive function is progressive.
For INRs as defined in \eqref{eq:inr}, this yields the following corollary to \cref{thm:inr}.
\begin{corollary}\label{coro:prog}
  Let $\Gamma\subseteq\widehat{\mathbb{R}}^d$ be conic, and let $\psi:\mathbb{R}^d\to\mathbb{C}$ be given, with Fourier support denoted $\Gamma_0:=\supp{(\widehat{\psi})}$.
  Let $\mathbf{W}^{(0)}\mathbf{r}=[\mathbf{W}_1\mathbf{r},\ldots,\mathbf{W}_{F_1}\mathbf{r}]^\top$ for $\mathbf{W}_1,\ldots,\mathbf{W}_{F_1}\in\mathbb{R}^{d\times d}$ each having full rank.
  Assume that for each $t=1,\ldots,F_1$, we have $\mathbf{W}_t^\top\Gamma_0\subseteq\Gamma$.
  Then, the WIRE INR $f_{\boldsymbol{\theta}}:\mathbb{R}^d\to\mathbb{C}$ defined by \eqref{eq:inr} is a $\Gamma$-progressive function.

  Moreover, if we fix some $\mathbf{r}_0\in\mathbb{R}^d$, and if the assumption $\mathbf{W}_t^\top\Gamma_0\subseteq\Gamma$ holds for the indices $t$ such that $\mathbf{W}_t\mathbf{r}+\mathbf{b}_t\in\supp(\psi)$, then $f_{\boldsymbol{\theta}}$ is locally $\Gamma$-progressive at $\mathbf{r}_0$.
\end{corollary}
The proof is left to \cref{app:prog}.
\Cref{coro:prog} shows that INRs preserve the Fourier support of the transformed template functions in the first layer up to conic combinations, so that any advantages/limitations of approximating functions using such $\Gamma$-progressive functions are maintained.
\begin{remark}
    One may notice that a $\Gamma$-progressive function will always incur a large error when approximating a real-valued function, as real-valued functions have conjugate symmetric Fourier transforms (apart from the case $\Gamma=\widehat{R}^d$).
    For fitting real-valued functions, it is effective to simply fit the \emph{real part} of the INR output to the function, as taking the real part of a function is equivalent to taking half of the sum of that function and its conjugate mirror in the Fourier domain.
    In the particular case of $d=1$, fitting the real part of a progressive INR to a function is equivalent to fitting the INR to that function's Hilbert transform.
\end{remark}

\subsection{Band-pass Progressive Wavelets}
\label{sec:complex:bp}

\begin{figure*}
  \centering
  \resizebox{0.9\linewidth}{!}{\input{figs/cones/cones.tikz}}
  \caption{(Left) A conic set $\Gamma_1\subset\mathbb{R}^2$, and a weakly conic set $\Gamma_2\subset\mathbb{R}^2$ (both truncated for illustration purposes). (Center) Modulus of a function $g$ given by the sum of four template atoms. (Right) Fourier transform of $f_{\boldsymbol{\theta}}(\mathbf{r})=\rho(g(\mathbf{r}))$, where $\rho(z)=-z+z^2-z^3$. The blue and orange cones correspond to the respectively highlighted parts of the function $g$. Effective Fourier supports of the template atoms constituting $g$ are enclosed by rectangles, and approximate centers of Frequency support for each atom and product of atoms are marked by colored circles.}
  \label{fig:conic}
\end{figure*}
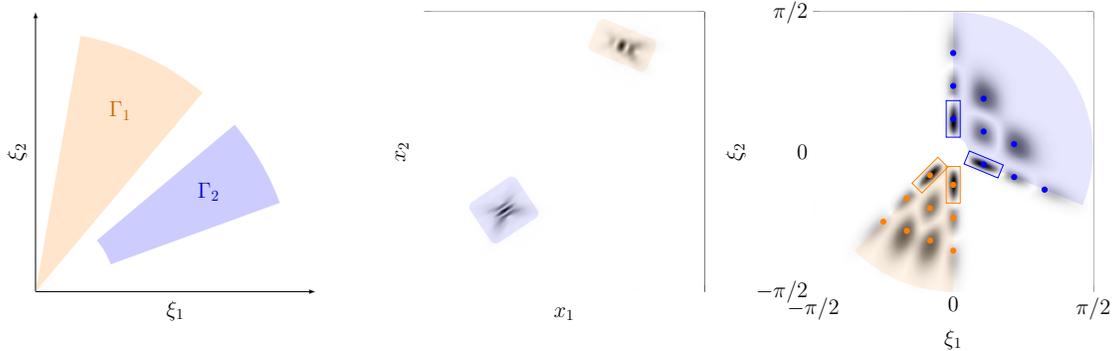

\Cref{coro:prog} holds for conic sets $\Gamma$, but is also true for a larger class of sets.
If some set $\Gamma\subseteq\mathbb{R}^d$ is conic, it is by definition closed under all sums with nonnegative coefficients.
Alternatively, consider the following weaker property:
\begin{definition}\label{defn:weakly-conic}
    Let $\Gamma\subseteq\widehat{\mathbb{R}}^d$.
    $\Gamma$ is said to be \emph{weakly conic} if for all $\gamma_1,\gamma_2\in\Gamma$ and $a_1,a_2\geq 1$, we have that $a_1\gamma_1+a_2\gamma_2\in\Gamma$, and that $0\in\Gamma$.
    A function $\psi:\mathbb{R}^d\to\mathbb{C}$ is said to be \emph{$\Gamma$-progressive} if $\supp(\widehat{\psi})\subseteq\Gamma$.
    The function $\psi$ is said to be \emph{locally $\Gamma$-progressive at $\mathbf{r}_0\in\mathbb{R}^d$} if there exists some $\Gamma$-progressive function $\psi_{\mathbf{r}_0}:\mathbb{R}^d\to\mathbb{C}$ so that for all smooth functions $\phi\in\mathcal{C}^\infty(\mathbb{R}^d)$ with support in a sufficiently small\footnote{Again, not necessarily compact.} neighborhood of $\mathbf{r}$, we have
    \begin{equation}
        \widehat{\phi\cdot\psi} = \widehat{\phi}\ast\widehat{\psi_{\mathbf{r}_0}}.
    \end{equation}
\end{definition}
The notion of a weakly conic set is illustrated in \cref{fig:conic}~(Left).
Just as in the case of progressive functions for a conic set, the set of $\Gamma$-progressive functions for a weakly conic set $\Gamma\subseteq\widehat{\mathbb{R}}^d$ constitutes an algebra over $\mathbb{C}$.
One can check, then, that \cref{coro:prog} holds for weakly conic sets as well.
Putting this into context, consider a template function $\psi$ such that $\widehat{\psi}$ vanishes in some neighborhood of the origin.
Assume furthermore that $\supp(\widehat{\psi})$ is contained in some weakly conic set $\Gamma$.

\begin{example}[Complex Meyer Wavelet]
    The complex Meyer wavelet is most easily defined in terms of its Fourier transform.
    Define
    \begin{equation*}
        \widehat{\psi}(\xi) := 
        \begin{cases}
            \sin(\frac{3\xi}{4}-\pi/2) & \xi\in[2\pi/3,4\pi/3] \\
            \cos(\frac{3\xi}{8}-\pi/2) & \xi\in[4\pi/3,8\pi/3] \\
            0 & \text{otherwise}.
        \end{cases}
    \end{equation*}
    The complex Meyer wavelet and its exponents are pictured in \cref{fig:stable}~(Right).
    Observe that these functions are not only progressive, but are also $\Gamma$-progressive for the weakly conic set $\Gamma=\hintCO{2\pi/3,\infty}$.
    The Meyer scaling function, pictured by the dashed line in \cref{fig:stable}~(Right), has Fourier support that only overlaps that of the complex Meyer wavelet, but none of its powers.
\end{example}

Applying this extension of \cref{coro:prog}, we see that if the atoms in the first layer of an INR using such a function $\psi$ have vanishing Fourier transform in some neighborhood of the origin, then the output of the INR has Fourier support that also vanishes in that neighborhood.
We illustrate this in $\mathbb{R}^2$ using a template function $\psi:\mathbb{R}^2\to\mathbb{C}$ where $\psi$ is the tensor product of a gaussian and a complex Meyer wavelet.
Using this template function, we construct an INR with $F_1=4$ in the first layer, and a single polynomial activation function.
The modulus of the sum of the template functions before applying the activation function is shown in \cref{fig:conic}~(Center).
We then plot the modulus of the Fourier transform of $f_{\boldsymbol{\theta}}$ in \cref{fig:conic}~(Right).
First, observe that since the effective supports of the transformed template functions are supported by two disjoint sets, the Fourier transform of $f_{\boldsymbol{\theta}}$ can be separated into two cones, each corresponding to a region in $\mathbb{R}^2$.
Second, since the complex Meyer wavelet vanishes in a neighborhood of the origin, these cones are weakly conic, so that the Fourier transform of $f_{\boldsymbol{\theta}}$ vanishes in a neighborhood of the origin as well, by \cref{coro:prog} applied to weakly conic sets.

\begin{remark}
  The weakly conic sets pictured in \cref{fig:conic}~(Right) are only approximation bounds of the true Fourier support of the constituent atoms.
  We see that \cref{coro:prog} still holds in an approximate sense, as the bulk of the Fourier support of the atoms is contained in each of the pictured cones.
\end{remark}

\subsection{A Split Architecture for INRs}
\label{sec:complex:split}

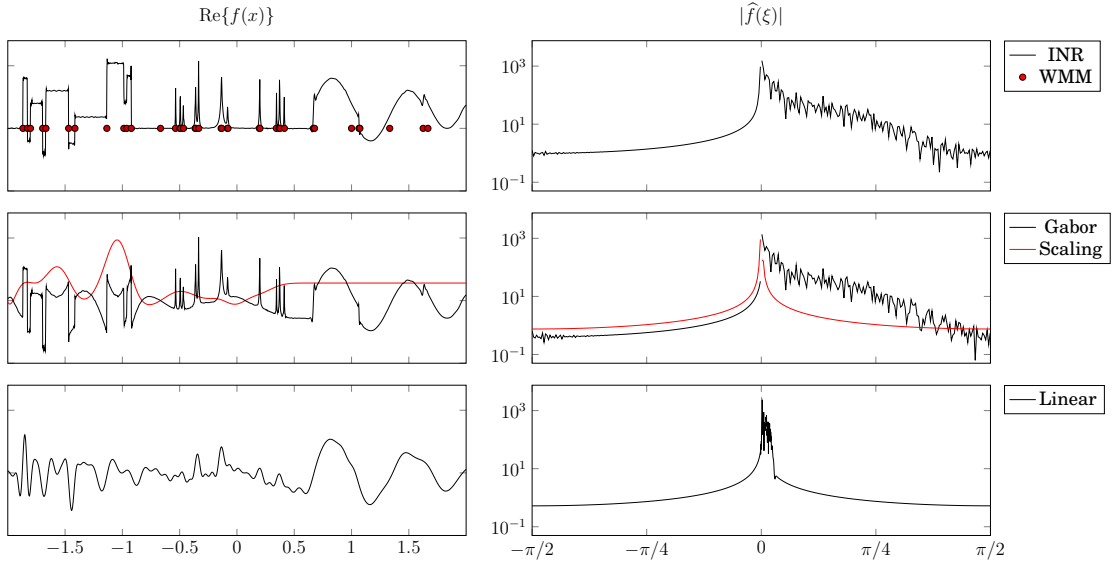
\begin{figure*}
  \centering
  \resizebox{0.9\linewidth}{!}{\input{figs/decouple/decouple.tikz}}
  \caption{(Left) Real part of INR evaluated on $\mathbb{R}$.
  (Right) Fourier transform of INR output.
  (Top) Sum of the Gabor INR and scaling INR, with wavelet modulus maxima points.
  (Center) Individual outputs of Gabor INR and scaling INR.
  (Bottom) Linear Gabor INR.}
  \label{fig:decouple}
\end{figure*}

Based on this property of INRs preserving the band-pass properties of progressive template functions, it is well-motivated to approximate functions using a sum of two INRs: one to handle the low-pass components using a scaling function, and the other to handle the high-pass components using a wavelet.
We illustrate this in \cref{fig:decouple}, where we fit two INRs to a test signal on $\mathbb{R}$~\citep{donoho1994}.

The first INR uses a gaussian template function $\psi(x)=\exp(-(\pi x)^2/6)$ with $L=1$, and the constraint that the weights $\mathbf{W}^{(0)}$ are all equal to one, \ie, the template atoms only vary in their abscissa.
Such a network is essentially a single-layer  perceptron~\citep{zhang1992} for representing smooth signals.
We refer to this network as the ``scaling INR.''

The second INR uses a Gabor template function $\psi(x)=\exp(-(\pi x)^2/6)\exp(-i2\pi x)$ with $L=3$, where we initialize the weights in the first layer to be positive, thus satisfying the condition of $\mathbf{W}_t^\top\Gamma\subseteq\Gamma$ in \cref{coro:prog} for $\Gamma=\mathbb{R}^+$.
Although $\psi$ is not progressive, its Fourier transform has fast decay, so we consider it to be essentially progressive, and thus approximately fulfilling the conditions of \cref{coro:prog}.
We refer to this network as the ``Gabor INR,'' as it is the WIRE architecture~\citep{saragadam2023} for signals on $\mathbb{R}$.

The reason for modeling a signal as the sum of a linear scaling INR and a nonlinear INR with a Gabor wavelet is apparent in \cref{fig:stable}~(Right), where the scaling function and powers of a complex Meyer wavelet are pictured.
Observe that the portions of the Fourier spectrum covered by the scaling function and the high powers of the Gabor wavelet (as in an INR, by \cref{thm:inr}) are essentially disjoint.
The idea behind this architecture is to use a simple network to approximate the smooth parts of the target signal, and then a more complicated nonlinear network to approximate the nonsmooth parts of the signal.

We plot the real part of the sum of the scaling INR and Gabor INR in \cref{fig:decouple}~(Top) along with the individual network outputs in \cref{fig:decouple}~(Center).
One can clearly see how the scaling INR captures the low-pass components of the signal, while the Gabor INR captures the transient behavior.

To see the role of the nonlinearities in the Gabor INR, we freeze the weights and biases in the first layer of the Gabor INR, and take an optimal linear combination of the resulting template atoms to fit the signal, thus yielding an INR with no nonlinearities beyond the template functions~\citep{zhang1992}.
The real part of the resulting function is plotted in \cref{fig:decouple}~(Bottom), where the singularities from the Gabor INR are severely smoothed.
This reflects how the activation functions resolve high-frequency features from low-frequency approximations, as illustrated initially in \cref{fig:conic}.

\section{Resolution of Singularities}
\label{sec:singular}

A useful model for studying sparse representations of images is the \emph{cartoon-like image}, which is a smooth function on $\mathbb{R}^2$ apart from singularities along a twice-differentiable curve~\citep{candes2004,wakin2006b}.

The smooth part of an image can be handled by the scaling function associated to a wavelet transform, while the singular parts are best captured by the wavelet function.
In the context of the proposed split INR architecture, the scaling INR yields a smooth approximation to the signal, and the Gabor INR resolves the remaining singularities.

\subsection{Initialization With the Wavelet Modulus Maxima}
\label{sec:singular:wmm}

As demonstrated by \cref{thm:inr}, the function $\psi$ in the first layer of an INR determines the expressivity of the network.
Many such networks satisfy a universal approximation property~\citep{zhang1992}, but their value in practice comes from their implicit bias~\citep{yuce2022,saragadam2023} in representing a particular class of functions.
For instance, using a wavelet in the first layer results in sharp resolution of edges with spatially compact error~\citep{saragadam2023}.
In the remainder of this section, we demonstrate how an understanding of singular points in terms of the wavelet transform can be used to bolster INR architectures and initialization schemes.

Roughly speaking, isolated singularities in a signal are points where the signal is nonsmooth, but is smooth in a punctured neighborhood around that point.
Such singularities generate ``wavelet modulus maxima'' (WMM) curves in the continuous wavelet transform~\citep{mallat1999}, which have slow decay in the Fourier domain.
With \cref{thm:inr} in mind, we see that INRs can use a collection of low-frequency template atoms and generate a collection of coupled high-frequency atoms, while also preserving the spatial locality of the template atoms.

The combination of these insights suggests a method for the initialization of INRs.
In particular, for a given number of template atoms $F_1$ in an INR, the network weights $\mathbf{W}^{(0)}$ and abscissa $\mathbf{b}^{(0)}$ should be initialized in a way that facilitates effective training of the INR via optimization methods.
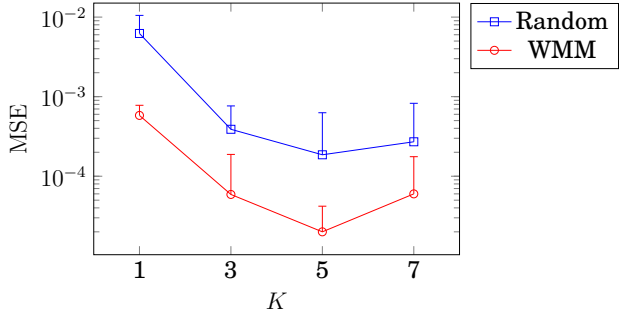
\begin{figure}
  \centering
  \resizebox{0.5\linewidth}{!}{\input{figs/wmm/initialize.tikz}}
  \caption{Comparison of initialization schemes. We plot the mean and standard deviation of the MSE over $10$ trials.}
  \label{fig:initialize}
\end{figure}
We empirically demonstrate the difference in performance for INRs initialized at random and INRs initialized in accordance with the singularities in the target signal.
Once again, we fit the sum of a scaling INR and a Gabor INR to the target signal in \cref{fig:decouple}.
In \cref{fig:initialize}, we plot the mean squared error (MSE) for this setting after $1000$ training steps for both randomly initialized and strategically initialized INRs, for $F_1=Km$, where $K\in\{1,3,5,7\}$ and $m$ is the number of WMM points as determined by an estimate of the continuous wavelet transform of the target signal.
The randomly initialized INRs have abscissa distributed uniformly at random over the domain of the signal.
The strategically initialized INRs place $K$ template atoms at each WMM point (so, a deterministic set of abscissa points).
Both initialization schemes randomly distribute the scale weights uniformly in the interval $[1,K]$.
We observe that for all $K$, the MSE of the strategically initialized INR is approximately an order of magnitude less than that of the randomly initialized INR.

\begin{figure*}
  \centering
  \resizebox{0.9\linewidth}{!}{\input{figs/wmm/2d/2d-initialize.tikz}}
  \caption{Image approximations and error using INRs that are (Left) randomly initialized and (Center) initialized using the Canny edge detector.
  (Right) PSNR of image approximations during training.}
  \label{fig:initialize2d}
\end{figure*}
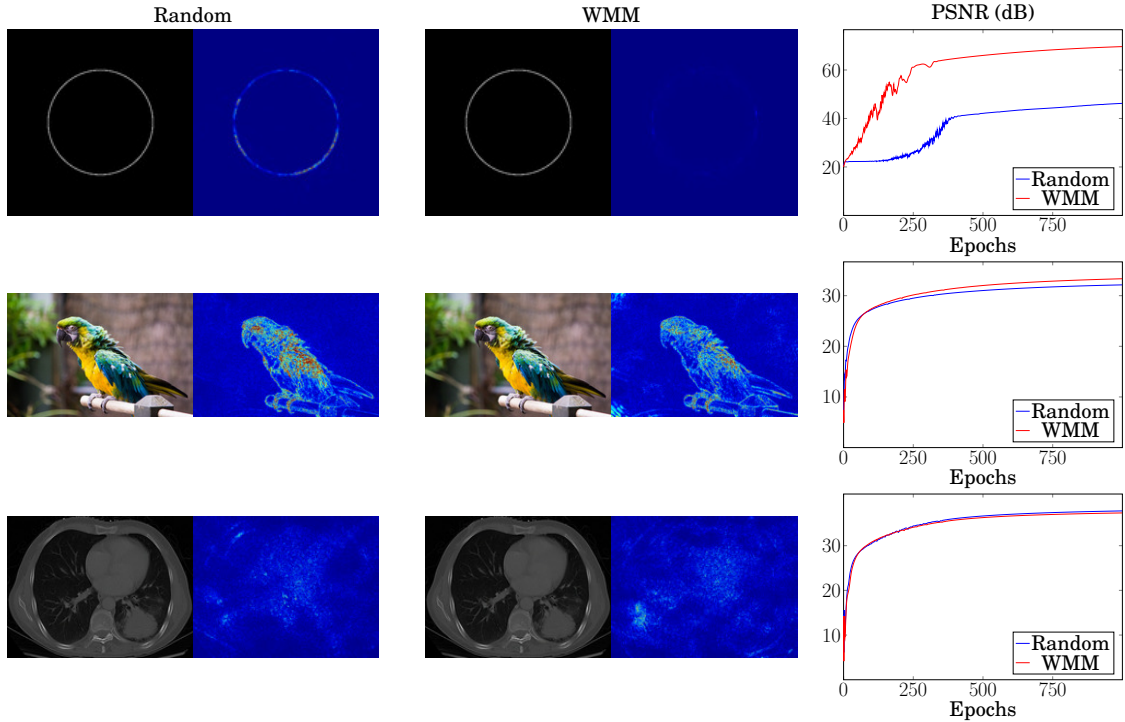
When $d=2$, \eg, images, the WMM can be approximated by the gradients of the target signal to obtain an initial set of weights and biases for the Gabor INR.
We evaluate this empirically on a set of images shown in \cref{fig:initialize2d}.
For each example, we approximate the target image using the proposed split INR architecture.
For the WMM-based initialization, we apply a Canny edge detector~\citep{canny1986computational} to encode the positions and directions of the edges.
Further details can be found in \cref{app:canny}.
We then used a subset of edge locations for biases of the first layer of the Gabor INR. For simplicty, we initialized the weights such that each neuron generates a radially symmetric Gabor filter.
For images consisting of isolated singularities, such as the circular edge example in the first row in \cref{fig:initialize2d}, we observe that a WMM-based initialization results in nearly $30$dB higher PSNR, along with faster convergence rates than its uninitialized counterpart.
A similar but less significant advantage can be seen for other images, such as the parrot with a blurred background (second row in \cref{fig:initialize2d}).
WMM-based initialization, has limited advantages when the image has dense texture, such as the chest X-ray scan (third row in \cref{fig:initialize2d}).
Overall, we observe that WMM-based initialization has similar advantages for signals on both $\mathbb{R}$ and $\mathbb{R}^2$.

\section{Conclusions}
\label{sec:conclusion}

We have offered a time-frequency analysis of INRs, leveraging polynomial approximations of the behavior of MLPs beyond the first layer.
By noting that progressive functions form an algebra over the complex numbers, we demonstrated that this analysis yields insights into the behavior of INRs using complex wavelets, such as WIRE~\citep{saragadam2023}.
This naturally leads to a split architecture for approximating signals, which decouples the low-pass and high-pass parts of a signal using two INRs, roughly corresponding to the scaling and wavelet functions of a wavelet transform.
Furthermore, the connection with the theory of wavelets yields a natural initialization scheme for the weights of an INR based on the wavelet modulus maxima of a signal.

INR architectures built using wavelet activation functions offer useful advantages for function approximation that balance locality in space and frequency.
The structure of complex wavelets as an algebra of functions with conic Fourier support, combined with the application of INRs for interpolating sampled functions, suggests a connection with \emph{microlocal and semiclassical analysis}~\citep{monard2023}.
As future work, we aim to extend the results of this paper to synthesize true singularities at arbitrarily fine scales, despite the continuous and non-singular structure of INR approximations of a sampled signal.
We also foresee the decoupling of the smooth and singular parts of a signal by the split INR architecture having useful properties for solving inverse problems and partial differential equations.

\section*{Acknowledgements}
\ifdefined\isfinal
This work was supported by NSF grants CCF-1911094, IIS-1838177, and IIS-1730574; ONR grants N00014-18-1-2571, N00014-20-1-2534, and MURI N00014-20-1-2787; AFOSR grant FA9550-22-1-0060; and a Vannevar Bush Faculty Fellowship, ONR grant N00014-18-1-2047.
Maarten de Hoop gratefully acknowledges support from the Department of Energy under grant DE-SC0020345, the Simons Foundation under the MATH+X program, and the corporate members of the Geo-Mathematical Imaging Group at Rice
University.
\else
DRAFT: acknowledgements redacted.
\fi

\newpage

\bibliographystyle{iclr2024_conference}
\bibliography{ref}

\appendix

\section{Proof of Theorem~\ref{thm:inr}}
\label{app:inr}

Let $\mathbf{r}_0$ be given, and let $I\subseteq\{1,\ldots,F_1\}$ be the set of indices such that $\mathbf{W}_t\mathbf{r}_0+\mathbf{b}_t\in\supp(\psi)$.
Then, there exists an open neighborhood $U\ni\mathbf{r}_0$ such that for all $\mathbf{r}\in U$, $\psi(\mathbf{W}_t\mathbf{r}+\mathbf{b}_t)=0$ for any $t\notin I$.
Thus, for any $\mathbf{r}\in U$, the assumptions on the activation functions imply that $f_{\boldsymbol{\theta}}(\mathbf{r})$ is expressible as a complex multivariate polynomial of $\{\psi(\mathbf{W}_i\mathbf{r}+\mathbf{b}_t)\}_{i\in I}$ with degree at most $K^{L-1}$, \ie{},
\begin{equation*}
    f_{\boldsymbol{\theta}}(\mathbf{r}) = \sum_{k=0}^{K^{L-1}}
    \sum_{\boldsymbol{\ell}\in\Delta(|I|,k)}
    \widehat{\beta}_{\boldsymbol{\ell}}\prod_{i\in I}\psi^{\ell_i}(\mathbf{W}_i\mathbf{r}+\mathbf{b}_i).
\end{equation*}
for some set of complex coefficients $\widehat{\beta}_{\boldsymbol{\ell}}$.
The desired result follows immediately from the convolution theorem.

\section{Relaxed Conditions on the INR Architecture}
\label{app:templates}

\Cref{thm:inr} assumes that the template function $\psi:\mathbb{R}^d\to\mathbb{C}$ has a Fourier transform that exists in the sense of tempered distributions, and that the activation functions are polynomials, which is a stronger condition than merely assuming they are complex analytic.
Here, we discuss ways in which this can be relaxed to include more general activation functions, as well as how template functions on Euclidean spaces other than $\mathbb{R}^d$ can be used.

\subsection{Relaxing the Class of Activation Functions}

In \cref{thm:inr}, two assumptions are made.
The first is that the template function has a Fourier transform that exists, possibly in the sense of tempered distributions.
The second is that the activation functions are polynomials of finite degree.
However, given that \cref{thm:inr} is a \emph{local} result, mild assumptions on the template function allow for reasonable extension of this result to analytic activation functions.

Indeed, if $\psi$ is assumed to be continuous, then one can take $V\subset U$ to be a compact subset of the neighborhood guaranteed by \cref{thm:inr}.
It follows, then, that the functions $\{\psi(\mathbf{W}_t\mathbf{r}+\mathbf{b}_t)\}_{t=1}^{F_1}$ are bounded over $V$.
Then, if the activation functions are merely assumed to be analytic, then the INR can be approximated uniformly well over $V$ by finite polynomials.
Without repeating the details of the proof, this yields an ``infinite-degree'' version of \cref{thm:inr}, where for any $\phi\in\mathcal{C}_0^\infty(V)$, we have
\begin{equation*}
    \widehat{\phi\cdot f_{\boldsymbol{\theta}}}(\xi) 
    = 
    \left( \sum_{k=0}^{\infty} \sum_{\boldsymbol{\ell}\in\Delta(F_1,k)} \widehat{\beta}_{\boldsymbol{\ell}} \left[ \widehat{\phi} \ast \bigast_{t=1}^{F_1} \left( e^{i2\pi\langle\mathbf{W}_t^{-\top}\xi,\mathbf{b}_t\rangle} \widehat{\psi}(\mathbf{W}_t^{-\top}\xi) \right)^{\ast \ell_t, \xi} \right] \right) (\xi).
\end{equation*}
This condition is not strong enough to handle general continuous activation functions, since the Stone-Weierstrass theorem for approximating complex continuous functions on a compact Hausdorff space requires polynomials terms to include conjugates of the arguments.
One could conceivably extend \cref{thm:inr} in this way, but this would not be compatible with the algebra of $\Gamma$-progressive functions, since $\Gamma$-progressive functions are not generally closed under complex conjugation.

\subsection{Template Functions From Another Dimension}

It is possible to consider cases where $\psi$ is defined as a map from a space of different dimension, say $\psi:\mathbb{R}^q\to\mathbb{C}$.
If $q<d$, then one can construct a map $\tilde{\psi}:\mathbb{R}^d\to\mathbb{C}$ so that $\tilde{\psi}(x_1,\ldots,x_q,x_{q+1},x_d)=\psi(x_1,\ldots,x_q)$.
This, for instance, is the case with SIREN~\citep{sitzmann2020} and the 1D variant of WIRE~\citep{saragadam2023}, where $\psi:\mathbb{R}\to\mathbb{R}$ is used for functions on higher-dimensional spaces.
In this case, the Fourier transform of $\tilde{\psi}$ only exists in the sense of distributions.
If $q>d$, then one can apply the results in this paper by treating the INR as a map from $\mathbb{R}^q\to\mathbb{C}$ followed by a restriction that ``zeroes out'' the excess coordinates by restricting the domain of $f_{\boldsymbol{\theta}}$ to the subspace of points $\mathbf{x}$ such that $\mathbf{x}=(x_1,\ldots,x_d,0,\ldots,0)$.

\section{Proof of Corollary~\ref{coro:prog}}
\label{app:prog}

We will show that the property of $f_{\boldsymbol{\theta}}$ being locally $\Gamma$-progressive at a point holds, as the result for being ``globally'' $\Gamma$-progressive follows.

Let $\mathbf{r}_0$ be given, and let $I\subseteq\{1,\ldots,F_1\}$ be the set of indices such that $\mathbf{W}_t\mathbf{r}_0+\mathbf{b}_t\in\supp(\psi)$.
By \cref{thm:inr}, there exists an open neighborhood $U\ni\mathbf{r}_0$ such that for all $\mathbf{r}\in U$, $\mathbf{f}_{\boldsymbol{\theta}}(\mathbf{r})$ takes the form of a complex multivariate polynomial of $\{\psi(\mathbf{W}_i\mathbf{r}+\mathbf{b}_t)\}_{i\in I}$.
Denoting this polynomial by $\mathcal{P}(\mathbf{r})$, we have that for any $\phi\in\mathcal{C}^\infty(U)$,
\begin{equation*}
    \phi\cdot\mathcal{P} = \phi\cdot f_{\boldsymbol{\theta}}.
\end{equation*}

Under the given assumptions, each of the terms $\{\psi(\mathbf{W}_i\mathbf{r}+\mathbf{b}_t)\}_{i\in I}$ is a $\Gamma$-progressive function.
Since $\Gamma$-progressive functions constitute an algebra over $\mathbb{C}$, any polynomial of them will yield a $\Gamma$-progressive function.
Letting $U$ be the ``sufficiently small neighborhood of $\mathbf{r}_0$,'' this implies that $f_{\boldsymbol{\theta}}$ is locally $\Gamma$-progressive at $\mathbf{r}_0$, as desired.

Noting that $\Gamma$-progressive functions also constitute an algebra over $\mathbb{C}$ when $\Gamma$ is weakly conic, this result also applies in that case.

\section{Wavelet Modulus Maximus Initialization in Two Dimensions}
\label{app:canny}

\begin{figure}[tt!]
    \centering
    \includegraphics[width=\linewidth]{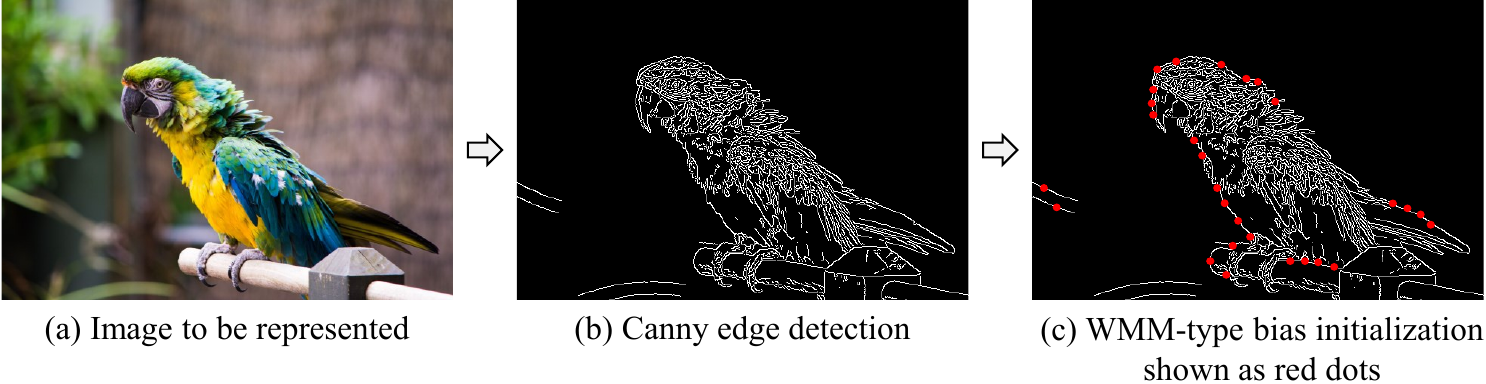}
    \caption{WMM initialization in two dimensions. (a) shows an image to be represented, while (b) shows the output of a Canny edge detector. (c) shows WMM-type bias initialization with the red dots showing some of locations used as bias values.}
    \label{fig:wmm2d}
\end{figure}

For 2D images, we approximate WMM location as edges of the image. \cref{fig:wmm2d} shows the flowchart for WMM-type initialization. We first start with the image to be represented and perform Canny edge dectection on it. We then use the binary edge map as a proxy for WMM locations. We then use the edge locations as bias values for each neuron in the Gabor INR. 

\end{document}

%% file: preamble.tex
\def\draftnotice{}

\ifdefined\isiclrpaper

    \usepackage{iclr2024_conference,times}
    \title{Implicit Neural Representations and the \\ Algebra of Complex Wavelets}
    \date{}
    \author{{T.~Mitchell~Roddenberry, Vishwanath~Saragadam, Maarten~V.~de~Hoop, Richard~G.~Baraniuk} \\ Rice University \\ \texttt{\{mitch,vishwanath.saragadam,mvd2,richb\}@rice.edu}}

    \ifdefined\isfinal
        \iclrfinalcopy
    \fi

\else 

    \usepackage{newcent}
    \usepackage{fullpage}
    \usepackage{natbib}
    \setcitestyle{authoryear,round,citesep={;},aysep={,},yysep={;}}

    \title{Implicit Neural Representations and the \\ Algebra of Complex Wavelets}
    \date{}
    
    \usepackage{authblk}
    
    \author[1]{T.~Mitchell~Roddenberry}
    \author[1]{Vishwanath~Saragadam}
    \author[2]{\\ Maarten~V.~de~Hoop}
    \author[1]{Richard~G.~Baraniuk}
    \affil[1]{Department of Electrical and Computer Engineering, Rice University}
    \affil[2]{Department of Computational Applied Mathematics and Operations Research, Rice University}
    \ifdefined\isfinal
        \def\draftnotice{}
    \else
        \def\draftnotice{\begin{center}\Large{DRAFT: Do not distribute.}\end{center}}
    \fi

\fi

\usepackage[T1]{fontenc}
\usepackage{amsmath,amsfonts,amssymb,amsthm,mathtools,accents}
\usepackage{relsize}
\usepackage{algorithm,algorithmic}

\usepackage{etoolbox}

\usepackage{float}

\usepackage[inline]{enumitem}

\usepackage{booktabs}

\usepackage{hyperref}
\hypersetup{
  colorlinks = true,
  allcolors = blue,
}
\usepackage[capitalise,nameinlink]{cleveref}
\usepackage{cite}
\usepackage{doi}
\usepackage{url}

\crefformat{app}{#2Appendix #1#3}
\Crefformat{app}{#2Appendix #1#3}

\usepackage{tikz}
\usetikzlibrary{matrix,arrows,arrows.meta,
shapes,positioning,
spy,
fit,calc,
decorations.pathmorphing,
intersections,
patterns,patterns.meta}

\usepackage{pgfplots}
\usepackage{pgfplotstable}
\pgfplotsset{compat=1.18}
\usepgfplotslibrary{groupplots}
\usepgfplotslibrary{fillbetween}

\newtheorem{theorem}{Theorem}

\newtheorem{corollary}[theorem]{Corollary}

\newtheorem*{example*}{Example}
\newtheorem*{theorem*}{Theorem}

\theoremstyle{definition}
\newtheorem{example}{Example}
\newtheorem{definition}[theorem]{Definition}

\theoremstyle{remark}
\newtheorem{remark}[theorem]{Remark}

\DeclareMathOperator{\supp}{supp}

\DeclareMathOperator{\RealPart}{Re}

\def\Re{\RealPart}

\newcommand{\bigast}{\mathop{\Huge \mathlarger{\mathlarger{\ast}}}}

\newcommand{\hintCO}[1]{\ensuremath{[#1)}}

\newcommand{\ie}{\textit{i.e.}}
\newcommand{\eg}{\textit{e.g.}}

\newcounter{edits}

\makeatletter

\newcommand\mitch[1]{\refstepcounter{edits}\noindent{\textcolor{green!50!black}{[\textbf{mitch:} #1]}}\addcontentsline{edits}{subsection}{{\thesubsection~(mitch) #1}}\gdef\there@is@an@edit{}}

\newcommand\vishwa[1]{\refstepcounter{edits}\noindent{\textcolor{blue}{[\textbf{vishwa:} #1]}}\addcontentsline{edits}{subsection}{{\thesubsection~(vishwa) #1}}\gdef\there@is@an@edit{}}

\newcommand\richb[1]{\refstepcounter{edits}\noindent{\textcolor{orange!80!black}{[\textbf{richb:} #1]}}\addcontentsline{edits}{subsection}{{\thesubsection~(richb) #1}}\gdef\there@is@an@edit{}}

\newcommand\mvdh[1]{\refstepcounter{edits}\noindent{\textcolor{magenta!80!black}{[\textbf{maarten:} #1]}}\addcontentsline{edits}{subsection}{{\thesubsection~(maarten) #1}}\gdef\there@is@an@edit{}}

\newcommand\hilite[1]{\refstepcounter{edits}\textcolor{magenta!80!black}{#1}\addcontentsline{edits}{subsection}{{\thesubsection~(hilite) #1}}\gdef\there@is@an@edit{}}

\newcommand{\fix}{\refstepcounter{edits}\marginpar{FIX}\addcontentsline{edits}{subsection}{{\thesubsection~(FIX)}}\gdef\there@is@an@edit{}}
\newcommand{\new}{\refstepcounter{edits}\marginpar{NEW}\addcontentsline{edits}{subsection}{{\thesubsection~(NEW)}}\gdef\there@is@an@edit{}}

\AtEndDocument{\ifdefined\there@is@an@edit\label{edit:was:used:in:doc}\fi}
\newcommand{\conditionalLoE}{\@ifundefined{r@edit:was:used:in:doc}{}{\section*{List of Editorial Marks}\@starttoc{edits}}}%

\makeatother

\newcounter{todo}

\makeatletter

\newcommand\todo[1]{\refstepcounter{todo}{\noindent\textcolor{red}{[\textbf{To-Do:} #1]}}\addcontentsline{tod}{subsection}{[\thetodo]~\thesubsection~#1}\gdef\there@is@a@todo{}}%

\AtEndDocument{\ifdefined\there@is@a@todo\label{todo:was:used:in:doc}\fi}
\newcommand{\conditionalLoT}{\@ifundefined{r@todo:was:used:in:doc}{}{\section*{List of To-dos}\@starttoc{tod}}}%

\makeatother

%% file: figs/zoo/zoo.tikz.tex
\begin{tikzpicture}[>=latex]

  \def\figdir{figs/zoo}

  \tikzstyle{every node}=[font=\Huge]

  \begin{groupplot}[
    group style={
      group size=6 by 1,
      group name=myplots,
      horizontal sep=1cm,
    },
    y tick label style={
        /pgf/number format/.cd,
            fixed,
            fixed zerofill,
            precision=2,
        /tikz/.cd
    },
    x tick label style={
        /pgf/number format/.cd,
            fixed,
            fixed zerofill,
            precision=2,
        /tikz/.cd
    },
    ylabel near ticks,
    width=8cm, height=8cm,
    xmin=0, xmax=1,
    ymin=0, ymax=1,
    no marks,
    legend pos=north east,
    axis line style={draw=none},
    xtick=\empty,
    ytick=\empty,
    xticklabels={,,},
    yticklabels={,,},
    scale only axis
    ]

    \pgfplotsforeachungrouped \f/\t in {
    gt/{Ground Truth},
    posenc/{Positional Enc.},
    sine/{SIREN},
    gauss/{Gaussian},
    gabor/{WIRE},
    gabor2d/{WIRE (2D)}
    }
    {
        \nextgroupplot[
        title={\t},
        ]
        \addplot graphics[xmin=0,xmax=1,ymin=0,ymax=1] {\figdir/\f.png};
        \draw[draw=red] (axis cs: {210/320},{200/320}) rectangle (axis cs: {240/320},{230/320});
        \addplot graphics[xmin=0.6,xmax=1,ymin=0,ymax=0.4] {\figdir/\f_zoomed.png};
        \draw[draw=red, thick] (axis cs: 0.6,0) rectangle (axis cs: 1,0.4);
    };

  \end{groupplot}

\end{tikzpicture}

%% file: figs/stable/stable.tikz.tex
\begin{tikzpicture}

  \def\figdir{figs/stable}

  \tikzstyle{every node}=[font=\Large]

  \pgfmathdeclarefunction{gauss}{2}{%
  \pgfmathparse{1/(sqrt(2*pi))*exp(-((x-#1)^2)/(2*#2^2))}%
  }

  \pgfmathdeclarefunction{levy}{2}{%
  \pgfmathparse{sqrt(#2/(2*pi))*exp(-#2/(2*(x-#1)))/((x-#1)^(3/2))*(x>#1)*(#2)}%
  }

  \pgfplotstableset{col sep=comma}

  \begin{groupplot}[
    group style={
      group size=4 by 1,
      group name=myplots,
      horizontal sep=2.25cm,
      vertical sep=1.5cm,
    },
    y tick label style={
        /pgf/number format/.cd,
            fixed,
            fixed zerofill,
            precision=1,
        /tikz/.cd
    },
    x tick label style={
        /pgf/number format/.cd,
            fixed,
            fixed zerofill,
            precision=1,
        /tikz/.cd
    },
    ylabel near ticks,
    width=6cm, height=6cm
    ]
    
    \nextgroupplot[
    title={Complex Sinusoid},
    xlabel={$\xi$},
    ylabel={$\widehat{\psi^n}(\xi)$},
    xmin=0, xmax=8,
    ymin=0, ymax=1.1,
    legend pos=outer north east,
    ]
    \addplot+[ycomb] coordinates { (2, 1) };
    \addplot+[ycomb] coordinates { (4, 1) };
    \addplot+[ycomb] coordinates { (6, 1) };

    \nextgroupplot[
    title={Gaussian},
    xlabel={$\xi$},
    ylabel={$\widehat{\psi^n}(\xi)$},
    domain=-5:5,
    samples=256,
    xmin=-3.5, xmax=3.5,
    ymin=0, ymax=0.43,
    no marks,
    legend pos=north east,
    ]
    \addplot {gauss(0,1)};
    \addplot {gauss(0,sqrt(2))};
    \addplot {gauss(0,sqrt(3))};

    \nextgroupplot[
    title={Gabor Wavelet},
    xlabel={$\xi$},
    ylabel={$\widehat{\psi^n}(\xi)$},
    domain=-1:9,
    samples=256,
    xmin=0, xmax=9,
    ymin=0, ymax=0.43,
    no marks,
    legend pos=north east,
    ]
    \addplot {gauss(2,1)};
    \addplot {gauss(4,sqrt(2))};
    \addplot {gauss(6,sqrt(3))};
    
    \nextgroupplot[
    title={Complex Meyer Wavelet},
    xlabel={$\xi$},
    ylabel={$\widehat{\psi^n}(\xi)$},
    xmin=0, xmax=20,
    ymin=0, 
    no marks,
    legend pos=outer north east,
    ]
    \addplot+[dashed, color=black] table[x=x, y=s] {\figdir/meyer_powers.csv};
    \addlegendentry{Scaling};
    \pgfplotsset{cycle list shift=-1}
    \addplot+ table[x=x, y=m1] {\figdir/meyer_powers.csv};
    \addlegendentry{$n=1$};
    \addplot+ table[x=x, y=m2] {\figdir/meyer_powers.csv};
    \addlegendentry{$n=2$};
    \addplot+ table[x=x, y=m3] {\figdir/meyer_powers.csv};
    \addlegendentry{$n=3$};

  \end{groupplot}

  
\end{tikzpicture}

%% file: figs/cones/cones.tikz.tex
\begin{tikzpicture}[>=latex]

  \tikzset{
    partial ellipse/.style args={#1:#2:#3}{
        insert path={+ (#1:#3) arc (#1:#2:#3)}
    }
  }

  \def\figdir{figs/cones}

  \newcommand\rtheta[2]{{#1*cos(#2)},{#1*sin(#2)}}

  \tikzstyle{every node}=[font=\Large]

  \pgfplotstableset{col sep=comma}

  \begin{groupplot}[
    group style={
      group size=4 by 1,
      group name=myplots,
      horizontal sep=2.5cm,
    },
    y tick label style={
        /pgf/number format/.cd,
            fixed,
            fixed zerofill,
            precision=2,
        /tikz/.cd
    },
    x tick label style={
        /pgf/number format/.cd,
            fixed,
            fixed zerofill,
            precision=2,
        /tikz/.cd
    },
    ylabel near ticks,
    width=8cm, height=8cm
    ]
    
    \nextgroupplot[
    xlabel={$\xi_1$},
    ylabel={$\xi_2$},
    axis lines=left,
    axis line style={-{>[scale=10]}},
    xtick=\empty,
    ytick=\empty,
    xticklabels={,,},
    yticklabels={,,},
    xmin=0, xmax=3.5,
    ymin=0, ymax=3.5,
    ]
    \draw[draw=none, fill=orange, opacity=0.2] (0,0) [partial ellipse=50:80:3.25 and 3.25] -- (0,0) -- cycle;
    \node at (65:2.5) {\textcolor{orange!80!black}{$\Gamma_1$}};

    \draw[draw=none, fill=blue, opacity=0.2] (0,0) [partial ellipse=20:40:3.25 and 3.25] -- (0,0) -- cycle;
    \draw[draw=none, fill=white] (0,0) [partial ellipse=19:41:1 and 1] -- (0,0) -- cycle;
    \node at (30:2.5) {\textcolor{blue!80!black}{$\Gamma_2$}};


    \nextgroupplot[
    xlabel={$x_1$},
    ylabel={$x_2$},
    xmin=-1, xmax=1,
    ymin=-1, ymax=1,
    no marks,
    xticklabels={,,},
    yticklabels={,,}
    ]
    \addplot graphics[xmin=-1,xmax=1,ymin=-1,ymax=1] {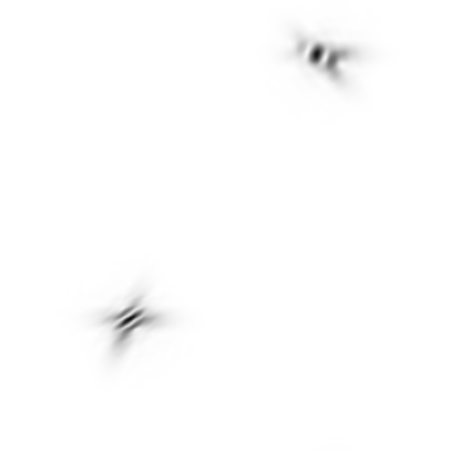};
    \draw[draw=none, rounded corners, fill=orange, opacity=0.1, rotate around={67.5:(0.1,0)}] 
    (axis cs: 0.8,-0.225) rectangle (axis cs: 1.05,0.225);
    \draw[draw=none, rounded corners, fill=blue, opacity=0.1, rotate around={-146.25:(0.1,0)}] 
    (axis cs: 0.575,-0.125) rectangle (axis cs: 0.975,0.225);

    \nextgroupplot[
    xlabel={$\xi_1$},
    ylabel={$\xi_2$},
    xmin=-20, xmax=20,
    ymin=-20, ymax=20,
    no marks,
    xtick = {-20,0,20},
    ytick = {-20,0,20},
    xticklabels={$-\pi/2$,$0$,$\pi/2$},
    yticklabels={$-\pi/2$,$0$,$\pi/2$}
    ]
    \addplot graphics[xmin=-20,xmax=20,ymin=-20,ymax=20] {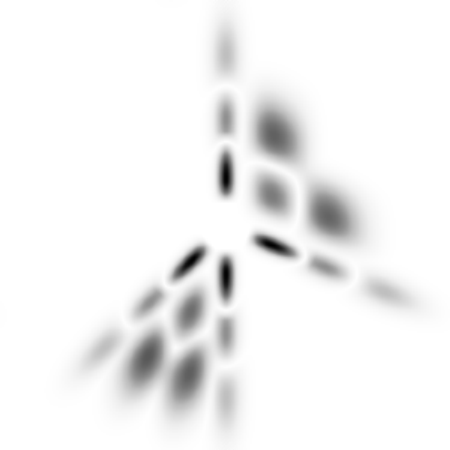};

    \draw[draw=none, fill=orange, opacity=0.1] (axis cs: 0,0) -- (axis cs: 0,-20) arc[start angle=270, end angle=225, radius=20] -- (axis cs:0,0) -- cycle;

    \draw[draw=none, fill=blue, opacity=0.1] (axis cs: 0,0) -- (axis cs: 0,20) arc[start angle=90, end angle=-22.5, radius=20] -- (axis cs:0,0) -- cycle;

    \draw[draw=none, fill=white] (axis cs: 0,0) circle ({2*pi/3});

    \draw[draw=orange, rotate around={-90:(0,0)}] (axis cs: {2*pi/3},-1) rectangle (axis cs: {7*pi/3},1);
    \draw[draw=orange, rotate around={-135:(0,0)}] (axis cs: {2*pi/3},-1) rectangle (axis cs: {7*pi/3},1);
    
    \draw[draw=none, fill=orange] (axis cs:\rtheta{4.5*pi/3}{-90}) circle (2pt);
    \draw[draw=none, fill=orange] (axis cs:\rtheta{4.5*pi/3}{-135}) circle (2pt);

    \draw[draw=none, fill=orange] (axis cs:\rtheta{9*pi/3}{-90}) circle (2pt);
    \draw[draw=none, fill=orange] (axis cs:\rtheta{1.85*4.5*pi/3}{-112.5}) circle (2pt);
    \draw[draw=none, fill=orange] (axis cs:\rtheta{9*pi/3}{-135}) circle (2pt);

    \draw[draw=none, fill=orange] (axis cs:\rtheta{13.5*pi/3}{-90}) circle (2pt);
    \draw[draw=none, fill=orange] (axis cs:\rtheta{2.78*4.5*pi/3}{-104.6}) circle (2pt);
    \draw[draw=none, fill=orange] (axis cs:\rtheta{2.78*4.5*pi/3}{-120.4}) circle (2pt);
    \draw[draw=none, fill=orange] (axis cs:\rtheta{13.5*pi/3}{-135}) circle (2pt);


    \draw[draw=blue, rotate around={-22.5:(0,0)}] (axis cs: {2*pi/3},-1) rectangle (axis cs: {7*pi/3},1);
    \draw[draw=blue, rotate around={90:(0,0)}] (axis cs: {2*pi/3},-1) rectangle (axis cs: {7*pi/3},1);
    
    \draw[draw=none, fill=blue] (axis cs:\rtheta{4.5*pi/3}{-22.5}) circle (2pt);
    \draw[draw=none, fill=blue] (axis cs:\rtheta{4.5*pi/3}{90}) circle (2pt);

    \draw[draw=none, fill=blue] (axis cs:\rtheta{9*pi/3}{-22.5}) circle (2pt);
    \draw[draw=none, fill=blue] (axis cs:\rtheta{1.11*4.5*pi/3}{33.75}) circle (2pt);
    \draw[draw=none, fill=blue] (axis cs:\rtheta{9*pi/3}{90}) circle (2pt);

    \draw[draw=none, fill=blue] (axis cs:\rtheta{13.5*pi/3}{-22.5}) circle (2pt);
    \draw[draw=none, fill=blue] (axis cs:\rtheta{1.86*4.5*pi/3}{7.24}) circle (2pt);
    \draw[draw=none, fill=blue] (axis cs:\rtheta{1.86*4.5*pi/3}{60.26}) circle (2pt);
    \draw[draw=none, fill=blue] (axis cs:\rtheta{13.5*pi/3}{90}) circle (2pt);



  \end{groupplot}

\end{tikzpicture}

%% file: figs/decouple/decouple.tikz.tex
\begin{tikzpicture}[>=latex, 
spy using outlines={rectangle, magnification=4, connect spies}]

  \def\figdir{figs/decouple}

  \tikzstyle{every node}=[font=\large]

  \pgfplotstableset{col sep=comma}

  \begin{groupplot}[
    group style={
      group size=2 by 3,
      group name=myplots,
      horizontal sep=1.5cm,
      vertical sep=0.5cm,
    },
    y tick label style={
        /pgf/number format/.cd,
            fixed,
            precision=0,
        /tikz/.cd
    },
    x tick label style={
        /pgf/number format/.cd,
            fixed,
            precision=1,
        /tikz/.cd
    },
    ylabel near ticks,
    width=12cm, height=5cm
    ]


    \nextgroupplot[
    title={$\Re\{f(x)\}$},
    no marks,
    xmin=-2.0, xmax=2,
    ymin=-5, ymax=7,
    yticklabels={,,},
    xticklabels={,,},
    xtick={-1.5,-1,-0.5,0,0.5,1,1.5},
    ]
    \addplot+[color=black] table[x=x, y expr={\thisrow{wavelet}+\thisrow{scaling}}] {\figdir/signals.csv};
    \addplot+[color=black, fill=red, only marks] table[x=x, y expr={0}] {\figdir/wmm.csv};
    
    \nextgroupplot[
    title={$|\widehat{f}(\xi)|$},
    legend pos=outer north east,
    no marks,
    xmin=-0.5, xmax=0.5,
    ymin=0.05, ymax=7500,
    xticklabels={,,},
    xtick={-0.5,-0.25,0,0.25,0.5},
    ymode=log,
    restrict y to domain=-5:10,
    each nth point=10,
    ]
    \addplot+[color=black] table[x=f, y=total] {\figdir/spectra.csv};
    \addlegendentry{INR};
    \addlegendimage{color=black, fill=red, only marks};
    \addlegendentry{WMM};

    \nextgroupplot[
    no marks,
    xmin=-2, xmax=2,
    ymin=-5, ymax=7,
    yticklabels={,,},
    xticklabels={,,},
    xtick={-1.5,-1,-0.5,0,0.5,1,1.5},
    ]
    \addplot+[color=red] table[x=x, y=scaling] {\figdir/signals.csv};
    \addplot+[color=black] table[x=x, y=wavelet] {\figdir/signals.csv};

    \nextgroupplot[
    legend pos=outer north east,
    no marks,
    xmin=-0.5, xmax=0.5,
    ymin=0.05, ymax=7500,
    xticklabels={,,},
    xtick={-0.5,-0.25,0,0.25,0.5},
    ymode=log,
    restrict y to domain=-5:10,
    each nth point=10,
    ]
    \addplot+[color=black] table[x=f, y=gabor] {\figdir/spectra.csv};
    \addlegendentry{Gabor};
    \addplot+[color=red] table[x=f, y=scaling] {\figdir/spectra.csv};
    \addlegendentry{Scaling};

    \nextgroupplot[
    no marks,
    xmin=-2, xmax=2,
    ymin=-5, ymax=7,
    yticklabels={,,},
    xtick={-1.5,-1,-0.5,0,0.5,1,1.5},
    ]
    \addplot+[color=black] table[x=x, y=linear] {\figdir/signals.csv};

    \nextgroupplot[
    legend pos=outer north east,
    no marks,
    xmin=-0.5, xmax=0.5,
    ymin=0.05, ymax=7500,
    xticklabels={$-\pi/2$,$-\pi/4$,$0$,$\pi/4$,$\pi/2$},
    xtick={-0.5,-0.25,0,0.25,0.5},
    ymode=log,
    restrict y to domain=-5:10,
    width=12cm,
    ]
    \addplot+[color=black] table[x=f, y=linear] {\figdir/spectra.csv};
    \addlegendentry{Linear};


  \end{groupplot}

\end{tikzpicture}

%% file: figs/wmm/initialize.tikz.tex
\begin{tikzpicture}[>=latex]

  \def\figdir{figs/wmm}

  \tikzstyle{every node}=[font=\large]

  \pgfplotstableset{col sep=comma}

  \pgfplotsset{style a/.style={color = blue, mark = square}}

  \pgfplotsset{style b/.style={color = red, mark = o}}

  \begin{axis}[
    y tick label style={
        /pgf/number format/.cd,
            fixed,
            precision=0,
        /tikz/.cd
    },
    xticklabels={1,3,5,7},
    xtick={1,3,5,7},
    xlabel={$K$},
    ylabel near ticks,
    ylabel={$\mathrm{MSE}$},
    width=6cm, height=6cm,
    legend pos=outer north east,
    xmin=0, xmax=8,
    ymin=0, ymax=0.015,
    ymode=log,
    width=8cm,
    ]

    \addplot+[style a, 
    error bars/.cd,
    y dir=plus, y explicit,
    ] coordinates {
    (1, 0.006226) +- (0, 0.004280)
    (3, 0.000389) +- (0, 0.000376)
    (5, 0.000186) +- (0, 0.000442)
    (7, 0.000271) +- (0, 0.000554)
    };

    \addlegendentry{Random};

    \addplot+[style b, 
    error bars/.cd,
    y dir=plus, y explicit,
    ] coordinates {
    (1, 0.000581) +- (0, 0.000198)
    (3, 0.000059) +- (0, 0.000129)
    (5, 0.000020) +- (0, 0.000022)
    (7, 0.000060) +- (0, 0.000116)
    };

    \addlegendentry{WMM};



  \end{axis}

\end{tikzpicture}

%% file: figs/wmm/2d/2d-initialize.tikz.tex
\begin{tikzpicture}[>=latex]

  \def\figdir{figs/wmm/2d}

  \tikzstyle{every node}=[font=\Huge]

  \begin{groupplot}[
    group style={
      group size=3 by 3,
      group name=myplots,
      horizontal sep=2cm,
      vertical sep=2cm
    },
    y tick label style={
        /pgf/number format/.cd,
            fixed,
            fixed zerofill,
            precision=0,
        /tikz/.cd
    },
    x tick label style={
        /pgf/number format/.cd,
            fixed,
            fixed zerofill,
            precision=0,
        /tikz/.cd
    },
    height=8cm,
    scale only axis
    ]
    
    \nextgroupplot[
    title={Random},
    xmin=0, xmax=2,
    ymin=0, ymax=1,
    no marks,
    legend pos=north east,
    axis line style={draw=none},
    xtick=\empty,
    ytick=\empty,
    xticklabels={,,},
    yticklabels={,,},
    width=16cm,
    ]
    \addplot graphics[xmin=0,xmax=1,ymin=0,ymax=1,includegraphics={clip}] {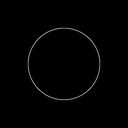};
    \addplot graphics[xmin=1,xmax=2,ymin=0,ymax=1] {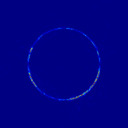};

    \nextgroupplot[
    title={WMM},
    xmin=0, xmax=2,
    ymin=0, ymax=1,
    no marks,
    legend pos=north east,
    axis line style={draw=none},
    xtick=\empty,
    ytick=\empty,
    xticklabels={,,},
    yticklabels={,,},
    width=16cm,
    ]
    \addplot graphics[xmin=0,xmax=1,ymin=0,ymax=1] {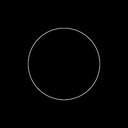};
    \addplot graphics[xmin=1,xmax=2,ymin=0,ymax=1] {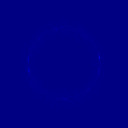};

    \nextgroupplot[
    title={PSNR (dB)},
    xlabel={Epochs},
    xmin=0, xmax=1000,
    ymin=0,
    no marks,
    legend pos=south east,
    xtick={0, 250, 500, 750},
    ytick={20, 40, 60},
    width=12cm,
    ]
    \addplot table[x expr=\coordindex, y index=0, header=false] {\figdir/circle_edge_psnr.csv};
    \addlegendentry{Random};
    \addplot table[x expr=\coordindex, y index=0, header=false] {\figdir/circle_edge_wmm_psnr.csv};
    \addlegendentry{WMM};


    \nextgroupplot[
    xmin=0, xmax=2,
    ymin=0, ymax=1,
    no marks,
    legend pos=north east,
    axis line style={draw=none},
    xtick=\empty,
    ytick=\empty,
    xticklabels={,,},
    yticklabels={,,},
    width=16cm,
    ]
    \addplot graphics[xmin=0,xmax=1,ymin=0.17,ymax=0.83] {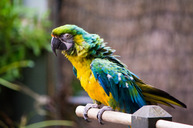};
    \addplot graphics[xmin=1,xmax=2,ymin=0.17,ymax=0.83] {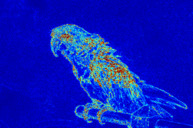};

    \nextgroupplot[
    xmin=0, xmax=2,
    ymin=0, ymax=1,
    no marks,
    legend pos=north east,
    axis line style={draw=none},
    xtick=\empty,
    ytick=\empty,
    xticklabels={,,},
    yticklabels={,,},
    width=16cm,
    ]
    \addplot graphics[xmin=0,xmax=1,ymin=0.17,ymax=0.83] {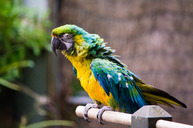};
    \addplot graphics[xmin=1,xmax=2,ymin=0.17,ymax=0.83] {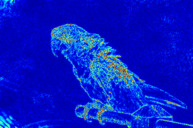};

    \nextgroupplot[
    xlabel={Epochs},
    xmin=0, xmax=1000,
    ymin=0,
    no marks,
    legend pos=south east,
    xtick={0, 250, 500, 750},
    ytick={10, 20, 30},
    width=12cm,
    ]
    \addplot table[x expr=\coordindex, y index=0, header=false] {\figdir/0886_psnr.csv};
    \addlegendentry{Random};
    \addplot table[x expr=\coordindex, y index=0, header=false] {\figdir/0886_wmm_psnr.csv};
    \addlegendentry{WMM};


    \nextgroupplot[
    xmin=0, xmax=2,
    ymin=0, ymax=1,
    no marks,
    legend pos=north east,
    axis line style={draw=none},
    xtick=\empty,
    ytick=\empty,
    xticklabels={,,},
    yticklabels={,,},
    width=16cm,
    ]
    \addplot graphics[xmin=0,xmax=1,ymin=0.12,ymax=0.88] {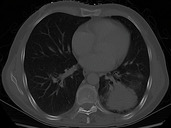};
    \addplot graphics[xmin=1,xmax=2,ymin=0.12,ymax=0.88] {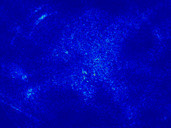};

    \nextgroupplot[
    xmin=0, xmax=2,
    ymin=0, ymax=1,
    no marks,
    legend pos=north east,
    axis line style={draw=none},
    xtick=\empty,
    ytick=\empty,
    xticklabels={,,},
    yticklabels={,,},
    width=16cm,
    ]
    \addplot graphics[xmin=0,xmax=1,ymin=0.12,ymax=0.88] {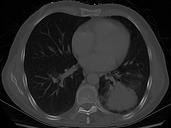};
    \addplot graphics[xmin=1,xmax=2,ymin=0.12,ymax=0.88] {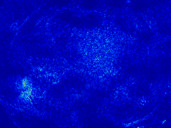};

    \nextgroupplot[
    xlabel={Epochs},
    xmin=0, xmax=1000,
    ymin=0,
    no marks,
    legend pos=south east,
    xtick={0, 250, 500, 750},
    ytick={10, 20, 30},
    width=12cm,
    ]
    \addplot table[x expr=\coordindex, y index=0, header=false] {\figdir/chest_psnr.csv};
    \addlegendentry{Random};
    \addplot table[x expr=\coordindex, y index=0, header=false] {\figdir/chest_wmm_psnr.csv};
    \addlegendentry{WMM};

  \end{groupplot}
  
\end{tikzpicture}

%% file: main.bbl
\begin{thebibliography}{17}
\providecommand{\natexlab}[1]{#1}
\providecommand{\url}[1]{\texttt{#1}}
\expandafter\ifx\csname urlstyle\endcsname\relax
  \providecommand{\doi}[1]{doi: #1}\else
  \providecommand{\doi}{doi: \begingroup \urlstyle{rm}\Url}\fi

\bibitem[Cand\`es(1998)]{candes1998}
Emmanuel~Jean Cand\`es.
\newblock \emph{Ridgelets: theory and applications}.
\newblock PhD thesis, Stanford University, 1998.

\bibitem[Cand{\`e}s \& Donoho(2004)Cand{\`e}s and Donoho]{candes2004}
Emmanuel~Jean Cand{\`e}s and David~Leigh Donoho.
\newblock New tight frames of curvelets and optimal representations of objects
  with piecewise $\mathcal{C}^2$ singularities.
\newblock \emph{Communications on Pure and Applied Mathematics}, 57\penalty0
  (2):\penalty0 219--266, 2004.

\bibitem[Canny(1986)]{canny1986computational}
John Canny.
\newblock A computational approach to edge detection.
\newblock \emph{IEEE Transactions on Pattern Analysis and Machine
  Intelligence}, \penalty0 (6):\penalty0 679--698, 1986.

\bibitem[Donoho \& Johnstone(1994)Donoho and Johnstone]{donoho1994}
David~Leigh Donoho and Iain~M Johnstone.
\newblock Ideal spatial adaptation by wavelet shrinkage.
\newblock \emph{Biometrika}, 81\penalty0 (3):\penalty0 425--455, 1994.

\bibitem[Fathony et~al.(2020)Fathony, Sahu, Willmott, and Kolter]{fathony2020}
Rizal Fathony, Anit~Kumar Sahu, Devin Willmott, and J~Zico Kolter.
\newblock Multiplicative filter networks.
\newblock In \emph{International Conference on Learning Representations}, 2020.

\bibitem[Mallat(1999)]{mallat1999}
St{\'e}phane Mallat.
\newblock \emph{A wavelet tour of signal processing}.
\newblock Elsevier, 1999.

\bibitem[Marar et~al.(1996)Marar, Carvalho~Filho, and Vasconcelos]{marar1996}
Joao~Fernando Marar, Edson~CB Carvalho~Filho, and Germano~C Vasconcelos.
\newblock Function approximation by polynomial wavelets generated from powers
  of sigmoids.
\newblock In \emph{Wavelet Applications III}, volume 2762, pp.\  365--374.
  SPIE, 1996.

\bibitem[Mildenhall et~al.(2020)Mildenhall, Srinivasan, Tancik, Barron,
  Ramamoorthi, and Ng]{mildenhall2020}
Ben Mildenhall, Pratul~P Srinivasan, Matthew Tancik, Jonathan~T Barron, Ravi
  Ramamoorthi, and Ren Ng.
\newblock {NeRF}: Representing scenes as neural radiance fields for view
  synthesis.
\newblock In \emph{IEEE European Conference on Computer Vision}, 2020.

\bibitem[Monard \& Stefanov(2023)Monard and Stefanov]{monard2023}
Fran{\c{c}}ois Monard and Plamen Stefanov.
\newblock Sampling the {X}-ray transform on simple surfaces.
\newblock \emph{SIAM Journal on Mathematical Analysis}, 55\penalty0
  (3):\penalty0 1707--1736, 2023.

\bibitem[Ramasinghe \& Lucey(2021)Ramasinghe and Lucey]{ramasinghe2021}
Sameera Ramasinghe and Simon Lucey.
\newblock Beyond periodicity: Towards a unifying framework for activations in
  coordinate-{MLP}s.
\newblock In \emph{IEEE European Conference on Computer Vision}, 2021.

\bibitem[Saragadam et~al.(2023)Saragadam, LeJeune, Tan, Balakrishnan,
  Veeraraghavan, and Baraniuk]{saragadam2023}
Vishwanath Saragadam, Daniel LeJeune, Jasper Tan, Guha Balakrishnan, Ashok
  Veeraraghavan, and Richard~G Baraniuk.
\newblock {WIRE}: Wavelet implicit neural representations.
\newblock In \emph{IEEE/CVF Computer Vision and Pattern Recognition
  Conference}, pp.\  18507--18516, 2023.

\bibitem[Sitzmann et~al.(2020)Sitzmann, Martel, Bergman, Lindell, and
  Wetzstein]{sitzmann2020}
Vincent Sitzmann, Julien Martel, Alexander Bergman, David Lindell, and Gordon
  Wetzstein.
\newblock Implicit neural representations with periodic activation functions.
\newblock \emph{Advances in Neural Information Processing Systems}, 2020.

\bibitem[Tancik et~al.(2020)Tancik, Srinivasan, Mildenhall, Fridovich-Keil,
  Raghavan, Singhal, Ramamoorthi, Barron, and Ng]{tancik2020}
Matthew Tancik, Pratul~P. Srinivasan, Ben Mildenhall, Sara Fridovich-Keil,
  Nithin Raghavan, Utkarsh Singhal, Ravi Ramamoorthi, Jonathan~T. Barron, and
  Ren Ng.
\newblock Fourier features let networks learn high frequency functions in low
  dimensional domains.
\newblock \emph{Advances in Neural Information Processing Systems}, 2020.

\bibitem[Wakin et~al.(2006)Wakin, Romberg, Choi, and Baraniuk]{wakin2006b}
Michael~B Wakin, Justin~K Romberg, Hyeokho Choi, and Richard~G Baraniuk.
\newblock Wavelet-domain approximation and compression of piecewise smooth
  images.
\newblock \emph{IEEE Transactions on Signal Processing}, 15\penalty0
  (5):\penalty0 1071--1087, 2006.

\bibitem[Xu et~al.(2022)Xu, Wang, Jiang, Fan, and Wang]{xu2022}
Dejia Xu, Peihao Wang, Yifan Jiang, Zhiwen Fan, and Zhangyang Wang.
\newblock Signal processing for implicit neural representations.
\newblock \emph{Advances in Neural Information Processing Systems}, 35, 2022.

\bibitem[Y{\"u}ce et~al.(2022)Y{\"u}ce, Ortiz-Jim{\'e}nez, Besbinar, and
  Frossard]{yuce2022}
Gizem Y{\"u}ce, Guillermo Ortiz-Jim{\'e}nez, Beril Besbinar, and Pascal
  Frossard.
\newblock A structured dictionary perspective on implicit neural
  representations.
\newblock In \emph{IEEE/CVF Computer Vision and Pattern Recognition
  Conference}, 2022.

\bibitem[Zhang \& Benveniste(1992)Zhang and Benveniste]{zhang1992}
Qinghua Zhang and Albert Benveniste.
\newblock Wavelet networks.
\newblock \emph{IEEE Transactions on Neural Networks}, 3\penalty0 (6):\penalty0
  889--898, 1992.

\end{thebibliography}
